\documentclass[aps,prl,showpacs,twocolumn]{revtex4}
\usepackage{amssymb}

\usepackage{amsmath}
\usepackage{graphicx}
\usepackage{dsfont}

\begin{document}

\title{Proximity induced superconductivity in monolayer CuO$_{2}$ on the
cuprate substrates}
\author{Guo-Yi Zhu$^{1}$, Fu-Chun Zhang$^{2,3}$, and Guang-Ming Zhang$^{1,4}$%
}
\affiliation{$^{1}$State Key Laboratory of Low-Dimensional Quantum Physics and Department
of Physics, Tsinghua University, Beijing 100084, China. \\
$^{2}$Department of Physics, Zhejing University, Hangzhou 310027, China.\\
$^{3}$Collaborative Innovation Center of Advanced Microstructures, Nanjing
210093, China.\\
$^{4}$Collaborative Innovation Center of Quantum Matter, Beijing 100084,
China.}
\date{\today}

\begin{abstract}
To understand the recently observed high temperature superconductivity in
the monolayer CuO$_{2}$ grown on the Bi$_{{2}}${Sr}$_{{2}}${CaCu}$_{{2}}${O}$%
_{{8+}\delta }$ substrates, we propose a two band model of the hybridized
oxygen p$_{x}$ and p$_{y}$ orbitals with the proximity effect of the
substrate. We demonstrate that both the nodal and nodeless superconducting
states can be induced by the proximity effect, depending on the strengths of
the pairing parameters.
\end{abstract}

\maketitle

\section{Introduction}

High transition temperature (T$_{c}$) superconductivity in cuprates remains
one of the most challenging topics in condensed matter physics\cite%
{Muller,Anderson,AtoZ,LeeNagaosaWen}. Despite world-wide efforts in the past
30 years, the physics community has still not reached a consensus what
causes T$_{c}$ so high. All the high T$_{c}$ superconducting copper oxides
have layered structures, the superconducting layers CuO$_{{2}}$ are
sandwiched by non-conducting charge reservoir layers. Modulation of charge
carriers in the CuO$_{{2}}$ planes is realized through substitution of
chemical elements in non-conducting planes, a key parameter in study of the
high T$_{c}$ superconductors.

Recently, {Zhong, et. al.,{\ reported that a monolayer CuO}}$_{{2}}$ is
successfully grown {\ on Bi}$_{{2}}${Sr}$_{{2}}${CaCu}$_{{2}}${O}$_{{8+}%
\delta }${{\ (Bi-2212) substrates via molecular beam epitaxy (MBE)\cite{Xue}%
. Their result is interesting and important\cite{FCZhang}. Unlike the
sandwiched CuO}}$_{{2}}${\ layers in the bulk Bi-2212, a monolayer CuO}$_{{2}%
}${\ is on the BiO surface of the top substrates, which opens a new route
for most direct probe such as scanning tunneling microscopy (STM) on the
high T$_{c}$ copper oxides. }Two distinct and spatially separated energy
gaps are observed on the films: the V-shaped gap is similar to the gap
observed on BiO layer, and the U-shaped gap is of superconducting nature and
is nodeless. The latter is also immune to scattering by K, Cs, and Ag atoms.
The observed U-like gap is in striking contrast with the nodal gap in the d$%
_{x^{2}-y^{2}}$-wave pairing symmetry which is well established in the bulk
cuprates~\cite{Shen,Harlingen,Tsuei}. The reported superconductivity in the
monolayer CuO$_{2}$ raises two important questions. One is the nature of its
superconductivity: is it the same as the superconductivity of the Bi-2212
substrates, or a new superconducting state? The other is its pairing
symmetry.

We begin with a brief summary of the electronic structure of the high T$_{c}$
superconducting copper oxides. The parent compounds of the copper oxides are
anti-ferromagnetic Mott insulator, and superconductivity arises upon
chemical doping, which introduces charge carriers in the CuO$_{2}$ planes.
The layers in cuprates are generally charged, either with positive charge
such as in the BiO layer in Bi-2212 or with negative charge such as in the
sandwiched CuO$_{2}$ layer. The charge carriers on the CuO$_{2}$ plane in
the parent compound is $-2e$ per unit cell consisting of one copper and two
oxygen atoms. The copper has a valence of 2+ and is in 3d$^{9}$
configuration with a single hole of d$_{x^{2}-y^{2}}$ orbital, and oxygen
has a valence of 2- and is in configuration of 2p$^{6}$. Due to the strong
on-site Coulomb repulsion, each Cu-atom is occupied with a single 3d hole
carrying spin-1/2 moment. Chemical doping introduces additional holes into
CuO$_{2}$ plane. These additional holes primarily reside on the oxygen
sites, forming the Zhang-Rice spin singlets, which move through the square
lattice of Cu-ions by exchanging with neighboring Cu spin-1/2 moment. This
leads to an effective two-dimensional t-J model or large on-site repulsive
Hubbard model~\cite{ZhangRice}. This model describes some of elementary
low-temperature physics in hole doped cuprates. In the relevant parameter
region, the carrier density introduced by doping is typically $0.1\sim 0.25$
hole per unit cell in bulk superconducting cuprates.

We now turn to examine the electronic structure of the MBE grown monolayer
CuO$_{2}$ on the top of a BiO layer, which is the surface plane of charge
neutral Bi-2212 substrate. If we neglect the charge transfer from the
monolayer CuO$_{2}$ to the inner planes, the monolayer CuO$_{2}$ is charge
neutral as required by total charge neutrality. Therefore, the Cu ion has
valence of 2+, or 3d$^{9}$ configuration, while the oxygen-ion has valence
of 1- and is hence in configuration of 2p$^{5}$ in the monolayer CuO$_{2}$.
The charge carriers in the CuO$_{2}$ monolayer thus have additional\textit{\
}two holes in average on the oxygen-ions per unit cell, which is in contrast
with the bulk superconducting cuprates. It is expected that some charge
carriers on the monolayer CuO$_{2}$ may be transferred to the inner planes
of the substrate Bi-2212, so that the actual charge carriers on the
oxygen-ions will be slightly reduced. With such a large charge carrier
concentration, we expect the CuO$_{2}$ monolayer itself to be a good metal,
whose low energy physics is totally different from the cuprates near a Mott
insulator.

In this paper we propose that the superconductivity observed in the CuO$_{2}$
monolayer is proximity induced superconductivity from the substrate cuprate.
The primary reason in support of this scenario is that the transition
temperature of the superconductivity in the monolayer is essentially the
same as that of the Bi-2212 substrates as reported in Ref.\cite{Xue}. The
challenge to this scenario is to explain the U-shaped superconducting gap in
the monolayer. We shall examine a two-band model for the monolayer, which is
coupled to the d-wave superconducting substrate by proximity effect. We show
that in certain parameter region, a two-band d-wave proximity induced
superconductivity may be gapful with U-shaped gap. We expect the monolayer
to exhibit a lower T$_{c}$ if the substrate has a lower T$_{c}$ or
non-superconducting if the substrate is non-superconduct. So the present
theory can be tested and distinguished in experiments.

The rest part of the paper is organized as follows. In the next section, we
present a model Hamiltonian for proximity induced superconductivity in
monolayer CuO$_{2}$. In the third section, we discuss the possible nodeless
superconducting gap of the model and present numerical results for the phase
diagram on the nodal or nodeless superconducting phases. The paper ends with
a short summary and discussion.

\section{Two-band model for monolayer CuO$_{2}$ and its proximity effect
induced d-wave superconductivity}

In this section, we propose a two-band model for the monolayer CuO$_{2}$ on
the Bi-2212 substrate and examine the proximity effect induced d-wave
superconductivity. We first discuss the Hamiltonian part without the
pairing, i.e., a non-interacting electron system on a square lattice with
oxygen orbitals of either 2p$_{x}$ or 2p$_{y}$ as shown in Fig. 1, whose
Hamiltonian is given by
\begin{equation}
H_{0}=\sum_{\mathbf{k}\sigma }\left(
\begin{array}{cc}
c_{1,\mathbf{k}\sigma }^{\dagger } & c_{2,\mathbf{k}\sigma }^{\dagger }%
\end{array}%
\right) \left(
\begin{array}{cc}
\epsilon _{x} & \epsilon _{xy} \\
\epsilon _{xy} & \epsilon _{y}%
\end{array}%
\right) \left(
\begin{array}{c}
c_{1,\mathbf{k}\sigma } \\
c_{2,\mathbf{k}\sigma }%
\end{array}%
\right) ,
\end{equation}%
where
\begin{eqnarray}
\epsilon _{x} &=&-2\left( t_{x}\text{cos}k_{x}+t_{y}\text{cos}k_{y}\right)
-\mu ,  \notag \\
\epsilon _{y} &=&-2\left( t_{y}\text{cos}k_{x}+t_{x}\text{cos}k_{y}\right)
-\mu ,  \notag \\
\epsilon _{xy} &=&-4t_{xy}\text{cos}\frac{k_{x}}{2}\text{cos}\frac{k_{y}}{2},
\end{eqnarray}%
and $t_{x}$, $t_{y}$, and $t_{xy}$ are assumed to be positive parameters,
and characterize the nearest neighbor intra- and inter-orbital hopping
terms, respectively. More accurate description for the monolayer would also
include Cu-3d$_{x^{2}-y^{2}}$ orbital. The holes on the O-site (2p$_{x}$ and
2p$_{y}$) are strongly coupled to the localized Cu-3d$_{x^{2}-y^{2}}$ spin,
and the system may be described by the Kondo lattice model with two
conduction bands\cite{GMZhang}. Roughly speaking, the holes on the Cu-site
lead to a renormalization of the two conduction bands of O-orbitals as in
the usual Kondo lattice problem. From this point of view, the
non-interacting two-band model described in Eq.(1) is a simplified model for
the monolayer with the understanding that the conduction bands are
renormalized ones\cite{GMZhang}.
\begin{figure}[t]
\includegraphics[width=8cm]{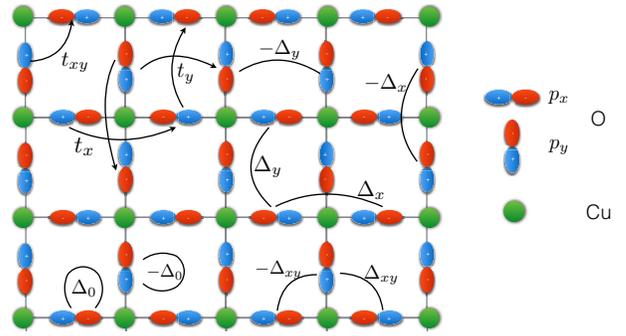} 
\caption{The monolayer CuO$_{2}$ on the cuprate Bi-2212 substrate and our
effective two-band model with all possible d-wave proximity pairings. }
\end{figure}

The inter-orbital hopping term hybridizes the two orbitals into two bands
with dispersion
\begin{equation}
\epsilon _{\pm }\left( \mathbf{k}\right) =\frac{\epsilon _{x}+\epsilon _{y}}{%
2}\pm \sqrt{\left( \frac{\epsilon _{x}-\epsilon _{y}}{2}\right)
{}^{2}+\epsilon _{xy}^{2}}.
\end{equation}%
The model is symmetric under reflection with respect to x-axis or y-axis
(Cu-O bonds), and has C$_{4}$ rotational symmetry:
\begin{eqnarray}
\epsilon _{x}(k_{x},k_{y}) &=&\epsilon _{y}(k_{y},-k_{x}),  \notag \\
\epsilon _{\pm }(k_{x},k_{y}) &=&\epsilon _{\pm }(k_{y},-k_{x}).  \notag
\end{eqnarray}

To describe the proximity effect of the cuprate substrate with d-wave
superconductivity, we introduce d-wave pairings for the carriers on the
oxygen 2p$_{x}$ and 2p$_{y}$ orbitals as shown in Fig.1. In the momentum
space, the full model Hamiltonian can be written as%
\begin{equation}
H=\sum_{\mathbf{k}}\Psi _{\mathbf{k}}^{\dagger }H\left( \mathbf{k}\right)
\Psi _{\mathbf{k}},
\end{equation}%
where the Nambu spinor has been introduced as $\Psi _{\mathbf{k}}^{\dagger
}=\left(
\begin{array}{cccc}
c_{1,\mathbf{k}\uparrow }^{\dagger } & c_{2,\mathbf{k}\uparrow }^{\dagger }
& c_{1,-\mathbf{k}\downarrow } & c_{2,-\mathbf{k}\downarrow }%
\end{array}%
\right) $, and the Hamiltonian matrix is given by%
\begin{equation}
H\left( \mathbf{k}\right) =\left(
\begin{array}{cccc}
\epsilon _{x} & \epsilon _{xy} & \Delta _{xx}(\mathbf{k}) & \Delta _{xy}(%
\mathbf{k}) \\
\epsilon _{xy} & \epsilon _{y} & \Delta _{xy}(\mathbf{k}) & \Delta _{yy}(%
\mathbf{k}) \\
\Delta _{xx}(\mathbf{k}) & \Delta _{xy}(\mathbf{k}) & -\epsilon _{x} &
-\epsilon _{xy} \\
\Delta _{xy}(\mathbf{k}) & \Delta _{yy}(\mathbf{k}) & -\epsilon _{xy} &
-\epsilon _{y}%
\end{array}%
\right) ,
\end{equation}%
with
\begin{eqnarray}
\Delta _{xx}\left( \mathbf{k}\right) &=&\Delta _{0}+2\left( \Delta _{y}\text{%
cos}k_{y}+\Delta _{x}\text{cos}k_{x}\right) ,  \notag \\
\Delta _{yy}\left( \mathbf{k}\right) &=&-\Delta _{0}-2\left( \Delta _{x}%
\text{cos}k_{y}+\Delta _{y}\text{cos}k_{x}\right) ,  \notag \\
\Delta _{xy}(\mathbf{k}) &=&4\Delta _{xy}\text{sin}\frac{k_{x}}{2}\text{sin}%
\frac{k_{y}}{2}.
\end{eqnarray}%
Note that the d-wave pairing symmetry of the substrates requires
\begin{equation*}
\Delta _{xx}\left( \mathbf{k}\right) =-\Delta _{yy}\left( \mathbf{\bar{k}}%
\right) ,\Delta _{xy}\left( \mathbf{k}\right) =-\Delta _{xy}\left( \mathbf{%
\bar{k}}\right) ,
\end{equation*}
where $\mathbf{\bar{k}=}\left( k_{y},-k_{x}\right) $. By transforming the
gap function matrix for $p_{x}$ and $p_{y}$ orbitals into the hybridized
band basis, we arrive at
\begin{equation*}
\tilde{\Delta}_{k}\left( \mathbf{k}\right) =\left(
\begin{array}{cc}
\Delta _{++}(\mathbf{k}) & \Delta _{+-}(\mathbf{k}) \\
\Delta _{+-}(\mathbf{k}) & \Delta _{--}(\mathbf{k})%
\end{array}%
\right) ,
\end{equation*}%
with
\begin{eqnarray*}
\Delta _{++}(\mathbf{k}) &=&\Delta _{d}(\mathbf{k})+\Delta _{s}(\mathbf{k})%
\text{cos}\theta _{\mathbf{k}}+\Delta _{xy}(\mathbf{k})\text{sin}\theta _{%
\mathbf{k}}, \\
\Delta _{--}(\mathbf{k}) &=&\Delta _{d}(\mathbf{k})-\Delta _{s}(\mathbf{k})%
\text{cos}\theta _{\mathbf{k}}-\Delta _{xy}(\mathbf{k})\text{sin}\theta _{%
\mathbf{k}}, \\
\Delta _{+-}(\mathbf{k}) &=&-\Delta _{s}(\mathbf{k})\text{sin}\theta _{%
\mathbf{k}}+\Delta _{xy}(\mathbf{k})\text{cos}\theta _{\mathbf{k}}, \\
\Delta _{s}(\mathbf{k}) &=&\Delta _{0}+\left( \Delta _{x}+\Delta _{y}\right)
\left( \cos k_{x}+\cos k_{y}\right) , \\
\Delta _{d}(\mathbf{k}) &=&\left( \Delta _{x}-\Delta _{y}\right) \left( \cos
k_{x}-\cos k_{y}\right) ,
\end{eqnarray*}%
where $\theta _{\mathbf{k}}=$tan$^{-1}\frac{2\epsilon _{xy}}{\epsilon
_{x}-\epsilon _{y}}$ in the range of $\left[ 0,\pi \right] $. It shows that
the hybridized bands of $p_{x}$ and $p_{y}$ orbitals are subjected to
mixture of s-wave and d-wave pairings together with $\Delta _{xy}(\mathbf{k}%
) $, which is neither s- or d-wave symmetry due to the mirror reflection
symmetry breaking with regards x- or y-axis. For consideration of mirror
symmetry, we will set $\Delta _{xy}=0$ in what follows. Actually an analysis
of the proximity pairing for the monolayer on top of the substrate Bi-2212
suggests the leading order in $\Delta _{xy}$ vanishes\cite{ChenWQ}. Neglect
of this term does not seem to change qualitative physics we wish to address.
Therefore the intra-band pairings $\Delta _{++}$ and $\Delta _{--}$ are
dominated by d-wave symmetry with a correction of modulated s-wave
component. Under the limit $\epsilon _{xy}>>|\epsilon _{x}-\epsilon _{y}|$,
we have $\theta _{\mathbf{k}}\rightarrow \pi /2$, the s-wave correction
vanishes, and the intra-band pairings are pure d-wave symmetric, which is
actually no surprise due to the proximity effect. However, the inter-band
pairing $\Delta _{+-}$ is pure s-wave symmetric, which plays a key role in
opening the nodal gap for d-wave pairing.

By diagonalizing the Hamiltonian matrix, we obtain the superconducting
quasiparticle spectrum. The quasiparticle dispersion takes the simple form
\begin{equation}
E_{\pm }\left( \mathbf{k}\right) =\sqrt{\frac{A_{\mathbf{k}}\pm \sqrt{A_{%
\mathbf{k}}^{2}-4B_{\mathbf{k}}}}{2}},
\end{equation}%
where
\begin{eqnarray}
A_{\mathbf{k}} &=&\epsilon _{x}^{2}+\epsilon _{y}^{2}+2\epsilon
_{xy}^{2}+\Delta _{xx}^{2}+\Delta _{yy}^{2},  \notag \\
B_{\mathbf{k}} &=&\left( \epsilon _{xy}^{2}-\epsilon _{x}\epsilon
_{y}\right) {}^{2}+2\Delta _{xx}\Delta _{yy}\epsilon _{xy}^{2}+\Delta
_{xx}^{2}\epsilon _{y}^{2}  \notag \\
&&+\Delta _{xx}^{2}\Delta _{yy}^{2}+\Delta _{yy}^{2}\epsilon _{x}^{2}.
\end{eqnarray}%
Two quasiparticle bands $E_{+}\left( \mathbf{k}\right) $ and $E_{-}\left(
\mathbf{k}\right) $ are separated, and $E_{-}\left( \mathbf{k}\right) $
corresponds to the lower one.

With the expression for the quasiparticle spectra, we are in good position
to examine the gap nodes and the STM probe of the gap. The d-wave pairing in
the bulk cuprates has the form of $\Delta _{d}\left( \mathbf{k}\right)
\varpropto (\cos {k_{x}}-\cos {k_{y})}$, and there are four nodes along the
lines of $k_{x}=\pm k_{y}$, which are crossing points of the lines with the
Fermi surface. As the intra-band pairing is dominated by d-wave pairing
form, it is expected that gap nodes in quasi-particle excitations are very
likely to occur. However, the mixture with s-wave pairing could potentially
open the gap nodes as long as the s-wave component is relatively strong
enough e.g. if $t_{xy}$ is small, and $\Delta _{0}$ is large. The exact
criterion is given by the zeroes of $B_{\mathbf{k}}$, which will be
discussed in the next section.

\section{Nodeless gap function in two-orbital model}

In this section, we examine the possibility of nodeless gap in the
two-orbital d-wave superconductivity in the previous section. From Eqs. (7),
(8), the condition for zeroes of the quasiparticle is given by $B_{\mathbf{k}%
}=0$. This depends on the hopping and pairing parameters. To illustrate the
possible nodeless gap, let us first consider a special case: $\epsilon _{xy}(%
\mathbf{k})=0$, namely the inter-orbital hopping vanishes. In this case, $B_{%
\mathbf{k}}=0$ requires both $\epsilon _{x}(\mathbf{k})=\epsilon _{y}(%
\mathbf{k})=0$ and $\Delta _{xx}(\mathbf{k})=0$ or $\Delta _{yy}(\mathbf{k}%
)=0$. These conditions cannot be satisfied in general except on some
discrete parameter space points. This simple example clearly demonstrates
the possible nodeless gap in the two-orbital d-wave superconductivity.
Actually this corresponds to the limit without hybridization, and the
intra-band pairing $\Delta _{xx}(\mathbf{k})$ and $\Delta _{yy}(\mathbf{k})$
are effectively $C_{4}$ anisotropic extended s-wave pairing. In passing, we
note that the d-wave gap function in a two orbital model allows a $\mathbf{k}
$-independent term in intra-orbital pairing, and the d-wave symmetry only
requires the opposite signs of this term for the two different orbitals, as
explicitly shown in Eq.(6).

Next, unlike the limit case we demonstrated above, we show that a weak
coupling pairing theory does not lead to a gapful d-wave state when $%
t_{xy}>>|t_{x}-t_{y}|$. This condition is more physically relevant since $%
t_{xy}$ is the nearest neighbor hopping. Within the weak coupling theory, we
may consider only intra-band pairing and neglect the inter-band pairing
since the two bands are not degenerate in the presence of the inter-orbital
hopping. Since then the intra-band pairing is almost pure d-wave symmetric,
the nodal structure in the d-wave pairing therefore remains in two-orbital
bands within the weak coupling theory. This shows that the vanishing of the
gap nodes requires a strong pairing coupling, comparable to the energy
splitting of the two bands. Below we shall consider a representative case to
illustrate the gap property. For the monolayer CuO$_{2}$, the band structure
parameters are not so known, and the proximity induced pairing strengths are
also not known. Therefore the choice of the parameters we will use below is
for the purpose of illustration.

We choose a set of parameters $t_{xy}=1$, $t_{x}=0.5$, $t_{y}=0.3$ for the
non-interacting part of the model. We will discuss the band structure and
then examine the gap property. The hybridized bands are plotted in Fig. 2.
\begin{figure}[t]
\includegraphics[width=8cm]{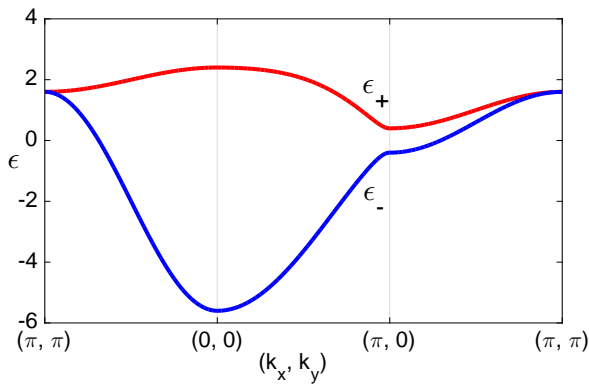} 
\caption{Two hybridized band dispersions for Hamiltonian $H_{0}$ in Eq. (1).
The parameters used are $t_{xy}=1$, $t_{x}=0.5$, $t_{y}=0.3$. The chemical
potential $\protect\mu $ is not included, or it is set equal to 0.}
\end{figure}
From Fig. 2, we can see that if $\mu <0.4$, only the lower hybridized band
is partially occupied and the higher band is completely empty. If $0.4<\mu
<1.6$, we have both the lower band and higher band partially occupied. For
further higher chemical potential, $\mu >1.6$, the lower band is fully
occupied. In this paper, we shall focus on the first case. The Fermi
surfaces for three typical values of $\mu $ within this region are plotted
in Fig. 3.
\begin{figure}[t]
\includegraphics[width=8cm]{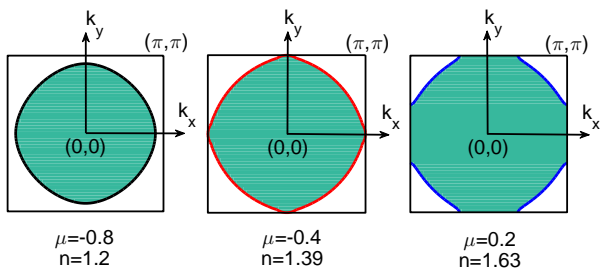} 
\caption{Fermi surfaces of model Hamiltonian $H_{0}$ in the region where the
lower energy band in Fig.2 is partially occupied and the higher energy band
is completely empty. The parameters for hopping integrals are the same as in
Fig.2. Three Fermi surfaces correspond to three different values of the
chemical potential.}
\end{figure}

In Fig. 4 we present a ''phase diagram'' for the nodal d-wave gapless and
nodeless d-wave gapful superconducting states in parameter space of $\Delta
_{0}$ and $\Delta _{x}$ and $\Delta _{y}$. It is interesting to note that
the gap property only depends on $\Delta _{x}+\Delta _{y}$ instead of each
of them in separate forms. At small values of $\Delta ^{\prime }s$, the gap
has nodes, consistent with the weak coupling pairing analyses. Note that $B_{%
\mathbf{k}}$ is non-negative, which is required for the quasiparticle
solutions as implied in Eq. (7). It can be seen that the nodeless phase only
occurs at large values of the on-site pairing strength $\Delta _{0}$.
Qualitatively, we may understand this result as that the gapless phase
requires strong inter-band pairing comparable to the band splitting.
\begin{figure}[t]
\includegraphics[width=7cm]{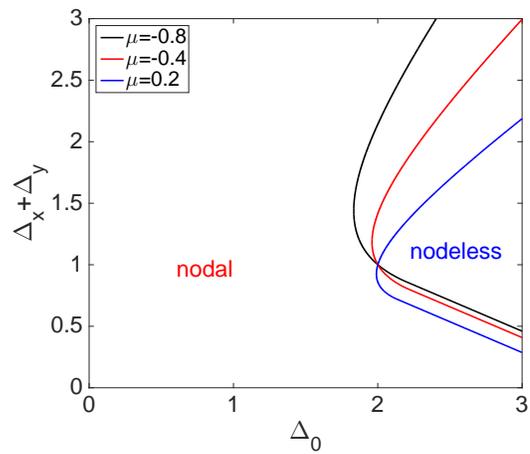} 
\caption{Phase diagram for the nodal and nodeless gap superconducting states
in the proximity induced two-band superconductivity. $\Delta _{0}$, $\Delta
_{x}$, and $\Delta _{y}$ are pairing amplitudes defined in Eq. (6). The
parameters in the non-interacting two-orbital model $H_{0}$ are: $t_{xy}=1$,
$t_{x}=0.5$, $t_{y}=0.3$, the same as in Fig. 2.}
\end{figure}

In the nodeless superconducting phase with $\Delta _{x}=0.6$, $\Delta
_{y}=0.3$, $\Delta _{0}=2.1$, and $\mu =0.2$, the lower quasiparticle band $%
E_{-}\left( \mathbf{k}\right) $ in the Brillouin zone is calculated and
displayed in Fig. 5a. This quasiparticle band has dramatic changes, very
different from the corresponding band without the pairing. The local density
of states, which is proportional to the local differential tunneling
conductance STM probes, can be calculated by using the quasiparticle
dispersion and is plotted in Fig.5b. Although there appears a U-shaped
mini-gap in the lower energy regime, the resonant peak does not reside
exactly on the edge of the mini-gap, rather different from the conventional
single band model, because the resonant peak still has its origin from
d-wave pairing. As a comparison, we also give rise to the results for the
nodal superconducting phase with $\Delta _{x}=0.6$, $\Delta _{y}=0.3$, $%
\Delta _{0}=0.9$, and $\mu =0.2$, and the corresponding results are
displayed in Fig.5c and 5d.
\begin{figure}[t]
\includegraphics[width=8cm]{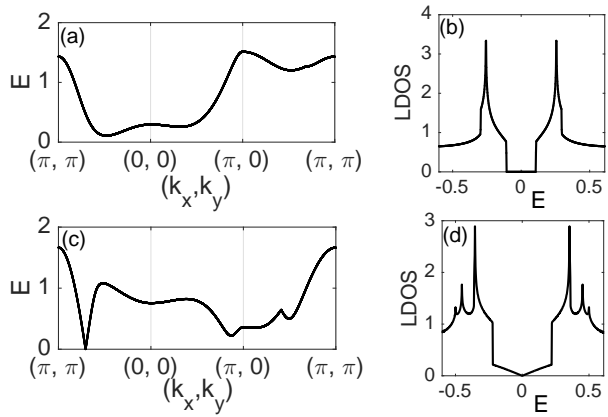} 
\caption{Upper panel: the lower quasiparticle dispersion $E_{-}(\mathbf{k})$
(a), and local density of states (b) in the nodeless supeconducting phase ($%
\Delta _{0}=2.1$). Lower panel: the lower quasiparticle band (c) and local
density of states (d) in the nodal gap phase ($\Delta _{0}=0.9$). All the
other parameters are the same in both phases: $\Delta _{x}=0.6$, $\Delta
_{y}=0.3$, and $\protect\mu =0.2$, and the non-interacting band structure
parameters are $t_{xy}=1$, $t_{x}=0.5$, and $t_{y}=0.3$, the same as in
Figs. 2-4.}
\end{figure}

\section{Summary and Discussions}

In summary, the successful growth of the monolayer CuO$_{2}$ on Bi-2212
reported by Zhong et al.~\cite{Xue} has provided a new material, as a
complement to bulk cuprates, to study physics in copper oxides. Motivated by
their new finding, we have proposed that the observed high $T_{c}$
superconductivity with nodeless gap is proximity induced superconductivity
and that the normal state of the monolayer is described by a two-orbital
model. We have further examined the superconducting gap functions in a
two-orbital model and demonstrated a mixture of d-wave and s-wave pairing,
which may explain the observed U-shaped gap in the experiment. In our
calculations, the nodeless gap phase in the two-orbital model occurs in the
region where the on-site pairing coupling is strong and comparable to the
energy splitting in the two bands.

We wish to point out that the non-interacting two orbital model we used in
the paper is a simplified model and the coupling between the O-2p bands and
the localized spin on the Cu sites has the Kondo coupling, which is expected
to greatly reduce the band widths of the O-orbitals near the Fermi level\cite%
{GMZhang}. From this point of view, the nodeless gap phase may be realized
in the monolayer CuO$_{2}$. While our theory is more closely related to the
monolayer CuO$_{2}$, our results may be relevant to nodeless d-wave
superconductivity in heavy fermion superconductor CeCu$_{2}$Si$_{2}$, where
the superconductivity is believed to have d-wave symmetry\cite{Steglich1},
recent specific heat data and superfluid density indicates a full gap in its
low-energy excitations\cite{Steglich2,Yuan}.

After finished this paper, we noted that nodeless excitation spectrum in a
two-orbital model with d-wave symmetry was previously discussed in the
context of iron based superconductivity\cite{Qimiao}. The reason for gapful 
excitations is due to the lack of intersection of the Fermi surfaces and the 
line nodes of the superconducting gap function, similar to the limiting case 
with vanishing inter-orbital hopping term we discussed in the beginning of 
the third section in this paper. More physically relevant case we discussed 
in this paper as illustrated in Fig. 2 to Fig.5 has a large inter-orbital
hopping integral, and the gapful excitation is resulted in the large
inter-band pairing.

\section{Acknowledgment}

We thank Qi-Kun Xue and his group members for stimulating discussions on
their experiments. We also thank W. Q. Chen for helpful discussions,
especially on the proximity induced pairing strength including the on-site
pairing. GMZ {acknowledges the support of NSF-China through Grant
No.20121302227}, and FCZ is supported in part by National Basic Research
Program of China (under grant No.2014CB921203) and NSFC (under grant
No.11274269).

\end{document}